\documentclass[aps,prl,twocolumn,amsmath,amssymb]{revtex4}
\usepackage{bm}
\usepackage{graphicx}

\newcommand{\be}{\begin{eqnarray}}
\newcommand{\ee}{\end{eqnarray}}

\newcommand{\ba}{\begin{array}}
\newcommand{\ea}{\end{array}}

\begin{document}

\title{Reply to ``Comment on `Security proof for cryptographic protocols based only on the monogamy of Bell's inequality violations' ''}
\author{M. Paw{\l}owski}
\affiliation{Department of Mathematics, University of Bristol, Bristol BS8 1TW, U.K.}
\begin{abstract}
In this reply I address the comment by W-Y. Hwang and O. Gittsovich on my paper [Phys. Rev. A {\bf 82}, 032313 (2010)]. The authors of the comment point out that I use implicit assumption that the alphabet of the eavesdropper is binary. They claim that such assumption is unrealistic. Here I show that even without this assumption the main result of my paper still holds.
\end{abstract}
\maketitle

The authors of the comment \cite{com} make an observation that
\be
P_B>P_E \Rightarrow I(\mathcal{A}:\mathcal{B})>I(\mathcal{A}:\mathcal{E})
\ee
holds only if the alphabet of $\mathcal{E}$ is binary. This is true as the example provided in the comment clearly shows. Since the proof presented in my paper \cite{pap} uses this implication in one step it holds only for the attacks of the adversary with the limited outcome alphabet, which is not very realistic. However, this problem can be easily overcome by the slight modification of the proof which I present below.

We may write the condition
\be \label{i0}
I(\mathcal{A}:\mathcal{B})>I(\mathcal{A}:\mathcal{E})
\ee
as
\be
H(\mathcal{A})-h(P_B)>H(\mathcal{A})-\sum_i p_i h(P(A=0|i))
\ee
where $h(.)$ is binary Shannon entropy function and $p_i$ is the probability that Eve's outcome is $i$. We may get rid of the identical terms on the both sides of the inequality and write it as
\be \label{i1}
h(P_B)<\sum_i p_i h(P(A=0|i)).
\ee
Let $P_{E|i}$ denote the probability of Eve correctly guessing the Alice's outcome when her outcome is $i$. Because $h(P(A=0|i))=h(P(A=1|i))$ and $P_{E|i}$ is just $\max\{P(A=0|i),P(A=1|i)\}$, we may substitute $h(P(A=0|i))=h(P_{E|i})$. $P_E$ is, obviously, equal to the weighted sum of these probabilities
\be
P_E=\sum_i p_i P_{E|i}.
\ee
Now, we can use the concavity of the binary entropy to claim that for the given value of $P_E$ the smallest value of $\sum_i p_i h(P_{E|i})$ is obtained when the probabilities $P_{E|i}$ take only the values $\frac{1}{2}$ or 1. If $p$ denote the sum of all $p_i$ such that $P_{E|i}=\frac{1}{2}$ then
\be
P_E=1-\frac{1}{2}p
\ee
and
\be
\sum_i p_i h(P_{E|i})\geq p=2(1-P_E).
\ee
Now, plugging it into (\ref{i1}), one sees that the sufficient condition to guarantee (\ref{i0}) is
\be
h(P_B)<2(1-P_E).
\ee
This is how Eq.(11) from my original paper should look like if the arbitrary alphabet for the adversary is allowed. Following the same line of reasoning as in the paper one can obtain the updated version of the general condition expressed in Eq.(27) which now becomes
\be \label{i2}
h(\beta_T(\mathcal{A},\mathcal{B}))<3-4f^M_{T'}(\beta_T(\mathcal{A},\mathcal{B})).
\ee

This new condition implies new critical values of $\beta(\mathcal{A},\mathcal{B})$ required for the security of the protocol. They can be found, just as in \cite{pap}, by substituting the monogamy condition for a given theory $T'$ for $f^M_{T'}(.)$ in (\ref{i2}). For QM monogamy the critical value is 0.841 which is below the Tsirelson bound (0.854). For NS monogamy the critical value is 0.881 which again is within the reach of no-signalling theories as they allow for the values up to 1 \cite{PR}.

Just as in \cite{pap} the condition (\ref{i2}) can be represented graphically and the critical values for different monogamies found as the coordinates of the points where the line corresponding to this condition intersects the one corresponding to the given monogamy. The updated version of the fig 1. from \cite{pap} is presented below.

\begin{figure}[ht]
\includegraphics[scale=0.85]{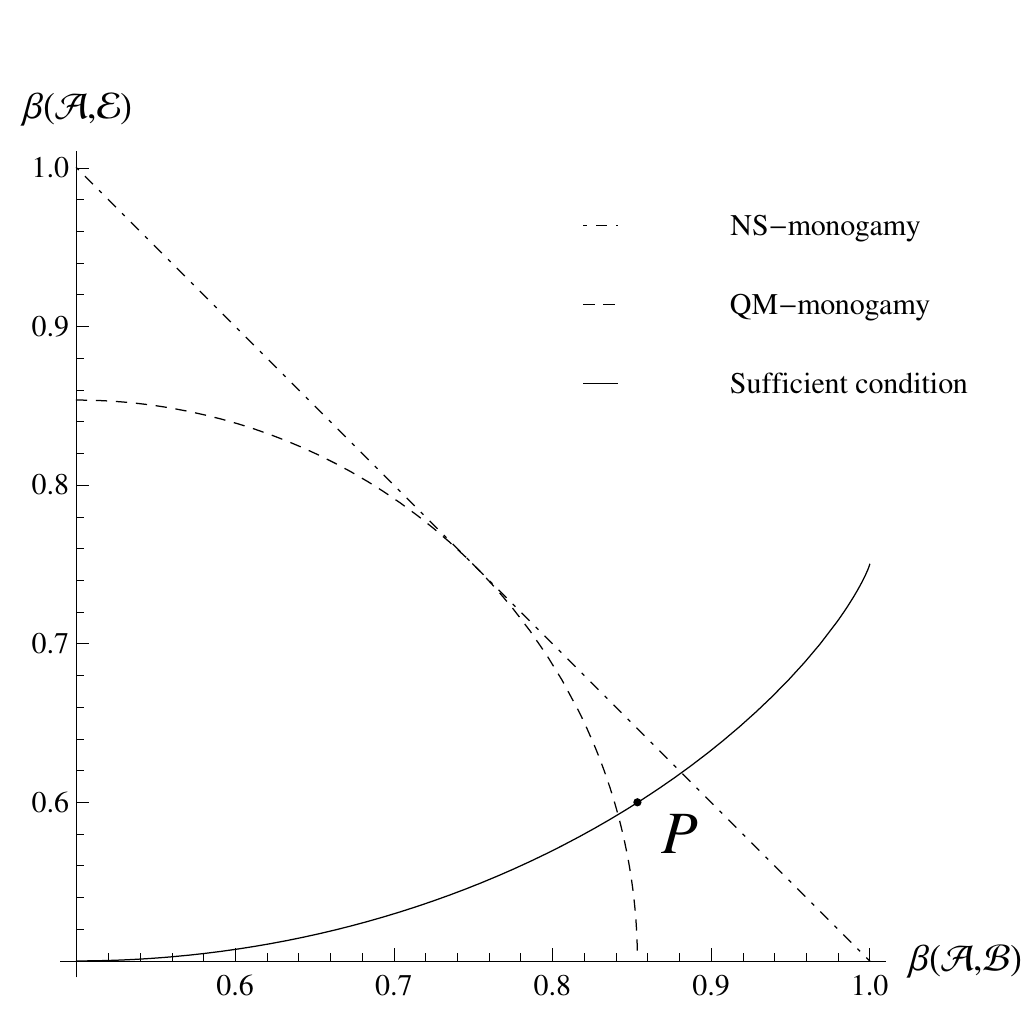}
\caption{Monogamy relations for different theories are plotted. The sufficient condition for the security of the key distribution protocol (\ref{i2}) is described by the line that passes through $P$.
The intersection of this line with monogamy relation for any specific theory T gives the critical value of $\beta_T(\mathcal{A},\mathcal{B})$ above which Alice and Bob can have secure communication. More explicitly, if $\mathcal{A}$ and $\mathcal{B}$ estimate their $\beta(\mathcal{A},\mathcal{B})$ to be greater than $\beta_T(\mathcal{A},\mathcal{B})$, they can have secure communication against individual attacks of the eavesdropper who has access to the resources of theory T.
Point $P$ has its horizontal coordinate equal to the Tsirelson bound, so for any theory T that intersects the sufficient condition line before $P$, it is possible for $\mathcal{A}$ and $\mathcal{B}$ to have a quantum protocol secure against attacks from T regime. With the unlimited alphabet of the eavesdropper QM monogamy still intersects the line corresponding to the condition \ref{i2} before $P$, which means that the quantum resources can guarantee the security against the eavesdropper limited by the monogamy resulting from the quantum theory. However, in contrast with the situation where the alphabet of the eavesdropper is binary, the same resources are not enough against the eavesdropper limited only by the no-signalling condition. 
}
\label{mons}
\end{figure}


\begin{thebibliography}{99}
\bibitem{pap} M. Paw{\l}owski, Phys. Rev. A {\bf 82}, 032313 (2010).
\bibitem{com} W-Y. Hwang, O. Gittsovich, comment to \cite{pap} accepted in PRA.
\bibitem{PR} S. Popescu, D. Rohrlich, Found. Phys. {\bf 24}, 379 (1994).
\end{thebibliography}
\end{document}